# Sound velocity measurements of tantalum under shock compression in the 10-110 GPa range

Jianbo Hu[1,a)], Chengda Dai[1,b)], Yuying Yu[1], Zijiang Liu[2], Ye Tan[1], Xianming Zhou[1], Hua Tan[1], Lingcang Cai[1], Qiang Wu[1]

[1] *Laboratory for Shock Wave and Detonation Physics Research, Institute of Fluid Physic Chinese Academy of Engineering Physics, Mianyang, Sichuan 621900, China*

[2] *Department of Physics, Lanzhou City University, Lanzhou 730070, China*

The high-pressure melting curve of tantalum (Ta) has been the center of a long-standing controversy. Sound velocities along the Hugoniot curve are expected to help in understanding this issue. To that end, we employed a direct-reverse impact technique and velocity interferometry to determine sound velocities of Ta under shock compression in the 10−110 GPa pressure range. The measured longitudinal sound velocities show an obvious kink at ~60 GPa as a function of shock pressure, while the bulk sound velocities show no discontinuity. Such observation could result from a structural transformation associated with a negligible volume change or an electronic topological transition.

## I. INTRODUCTION

Ever since Errandonea *et al.* performed the first measurement of the static high-pressure melting curve of tantalum (Ta) using a laser-heated diamond anvil cell (LHDAC), [1] controversy has surrounded the phase diagram of Ta. Melting temperatures at 300 GPa pressure obtained in shock-wave (SW) experiments (9700±1000 K) [2] and the extrapolation of LHDAC measurements (<4000 K) [1] show an astonishing discrepancy, but also great interest and intrigue as to its essence. In the last decade, a great deal of experimental [3-6] and theoretical [7-12] effort has been devoted to this issue and several possible explanations have been proposed; however, we have to admit that no agreement has been reached to date.

More recently, Dewaele *et al.* performed an exhaustive sequence of LHDAC measurements to extract the high-pressure melting curve of Ta. [6] They claimed that the flat curve observed previously, as well as those of other bcc transition metals, such as W, Mo, and V, could be induced by chemical reactions of metals with carbon and/or pressure-transmitting media, and by the pyrometer technique. After eliminating those effects, they observed more consistent results with SW experiments and most first-principle calculations. This observation is surely meaningful, but we also have to notice that in one LHDAC measurement Errandonea *et al*. have excluded the possibility of chemical reactions and pyrometer problems,[13] and used different measurement techniques to confirm their early observations.[3,4] Here we will not, and are also incapable to, judge which LHDAC measurement is more reliable, but just present our experimental results of high-pressure sound velocities along the Hugoniot that are expected to be extremely helpful in understanding the bcc-phase stability of Ta.[14] More than one theory has predicted that bcc-Ta will transform into other structures[8,11,12] under high pressure and high temperature prior to melting which might play an important role in solving the discrepancy in the high-pressure melting curves obtained from SW and LHDAC experiments.[15]

Although we only deal with Ta in this work, it is instructive to extend our discussion to other bcc transition metals, because these systematically show a flat melting tendency under static high pressure.[1]

## II. EXPERIMENTS

The reverse-impact geometry proposed by Duffy and Ahrens[16] was modified to obtain precise longitudinal sound velocities. The modified configuration is schematically illustrated in Fig.1. The impactor, *viz.* a Ta polycrystalline specimen (~2 mm) of 99.98% purity, was launched with an impact velocity, $W$, by either a two-stage light gas gun for higher-pressure experiments ($W \geq 2$ km/s) or a powder gun for lower-pressure ones ($W<2$ km/s). In contrast to the original reverse-impact geometry, a thin Al foil (~8 μm) instead of an Al-plate buffer was epoxy(~7 μm)-mounted on LiF single crystal with ~3 μm Al-film. The particle velocity history at the impact interface was measured by using a velocity interferometer system for any reflector (VISAR) with a time resolution of 1 ns. The modified reverse-impact geometry almost excluded the interruption from Al-plate buffer and enabled the VISAR to record a complete trace from loading to unloading. More experimental details and the analysis of wave interactions can be found in Ref. [17]. In this measurement geometry, the Lagrange longitudinal sound velocity, $C_L$, can be expressed as:

$$C_L = \frac{h_s}{\Delta t - h_s/D}, \qquad (1)$$

where $h_s$ is the thickness of the specimen, $D$ the shock wave velocity in a Ta impactor, and $\Delta t = t_B - t_A$. Here $t_A$ and $t_B$ are respectively the impact time and the time that the leading edge of rarefaction fan reaches the impact interface. Travel time of the shock wave in the Al foil and film has been neglected.

At lower shock pressures, if the elastic precursor is evident in the observed interfacial particle histories, $C_L$ is then modified:[16]

$$C_L = \frac{h_s - h_1}{\Delta t - (h_s + h_1)/C_{L_0}}, \qquad (2)$$

with

$$h_1 = h_s \frac{C_{L_0} - D}{C_{L_0} + D}, \qquad (3)$$

where $h_1$ is the thickness of the interaction region between the elastic precursor and shock wave at the rear of the sample, and $C_{L_0}$ the elastic precursor velocity taken from ultrasonic measurements.[18] The Euler longitudinal sound velocity, $C_l$, is obtained from:

$$C_l = (\rho_0 / \rho) C_L, \qquad (4)$$

where $\rho_0$ and $\rho$ are the respective densities at ambient conditions and at the Hugoniot state. This technique has been successfully applied to identify shock-induced structural transformations and melting of tin, and confirmed by static high-pressure experiments.[17]

## III. RESULTS AND DISCUSSION

Ten shots were performed to measure the interfacial particle histories at various impact velocities ranging from 0.989 km/s to 5.139 km/s. Fig.2 shows three wave profiles. At higher shock pressure (>31.3 GPa) where the elastic precursor is suppressed, $C_L$ is determined by Eq.(1). At lower shock pressure (≤31.3 GPa) where the elastic precursor is evident, $C_L$ is calculated from Eq.(2). For most of the velocity profiles obtained, the exact time of elastic-plastic transition during unloading is blurred by the Bauschinger effect. Thus the Lagrange bulk sound velocities, $C_B$, cannot be directly determined based on interfacial particle velocity (*u*) profiles. Asay *et al.* proposed a method by which to deduce $C_B$ at the Hugoniot state by linearly extrapolating the plastic unloading portion to the Hugoniot state in the plot of $C_L$ against engineering strain (*e*).[19] We noticed that, for the plastic segment, the $C_L(u)$ relation would offer better linearity than $C_L(e)$. Bezruchko *et al.* have also concluded that the plastic release wave velocity is approximately linear with particle velocity.[20] Therefore, the plastic unloading part of the $C_L(u)$ line was used to linearly extrapolate and obtain $C_B$ at the Hugoniot state. The uncertainty in $C_B$ is slightly larger than that in $C_L$ due to error propagation and wave interaction. The Euler bulk sound velocity, $C_b$, is calculated by a similar equation as Eq. (4). Plots of Euler velocities against shock pressure are presented in Fig.3.

The estimated longitudinal and bulk sound velocities, deduced from the Grüneisen equation of state and the assumption of $\rho\gamma = \rho_0\gamma_0$ =constant[21] (where *γ* is Grüneisen parameter,), are also shown in Fig.3. In combination with the sound velocity data reported in Ref.[2,10,22], we clearly see that there are two breaks in the $C_l$-plot at ~60 GPa and ~295 GPa. The discontinuity at ~295 GPa where $C_l$ collapsed onto $C_b$, has been well-acknowledged to result from shock-induced melting.[2] The other discontinuity at ~60 GPa appears for the first time. To our knowledge, there are two possible explanations for this discontinuity: a structural transformation or an electronic topological transition. We will discuss both in detail later in this section. It is also notable that there is no obvious break in the $C_b$ curve in Fig. 3. Therefore, we infer that no abrupt volume change takes place around those shock pressures at which the longitudinal sound velocity shows a break.

Usually, $C_l$ discontinuities are used to identify shock-induced polymorphous transformation and melting.[17, 23] Molybdenum, Mo, is a neighbor of Ta in the periodic table of elements that also exhibits controversial high-pressure melting behavior. High-pressure sound velocity measurements have identified a shock-induced solid-solid transition in Mo[24] that is further supported by first-principle calculations.[25,26] Note that disagreement also exists over the phase stability of bcc-Mo.[27] Thus, one possible interpretation of the observed kink at ~60 GPa is a shock-induced phase transition. Surprisingly, this interpretation is very consistent with a recent prediction by Burakovsky *et al.* using first-principle calculations.[12] They extensively studied the phase stability of bcc-Ta and predicted a possible phase transition from bcc to hexagonal omega phase (or other phase) at ~70 GPa (or <70 GPa). This transition is further supported from shock-recovery experiments, in which small pseudo-hexagonal omega particles were observed by using transmission electron microscopy in the shock-recovered polycrystalline Ta released from the Hugoniot state at 45 GPa.[28] Hsiung proposed a possible mechanism for the observed bcc to omega phase transformation, that is, the high shock pressure suppresses the dynamic-recovery reaction and leads to shear transformation in Ta.[29] The shear stress plays a crucial role, and is also the main difference between shock compression and hydrostatic-pressure compression in DAC. In the latter, no considerable shear deformation exists in the compressed specimen, thus no structural transformation has been observed up to 174 GPa.[30] It is worth noting that the yield strength, *Y*, of Ta shows a drastic drop at ~60

GPa as Ta undergoes inhomogeneous compression in DAC, significantly different than under homogeneous conditions.[31] This drop is in accordance with our present observation if we consider the change in shear modulus $G = 3\rho\left(C_l^2 - C_b^2\right)/4$ and the approximation of $Y/G \approx$ constant.[32] It has been postulated that no volume change is associated with the bcc-omega phase transition,[31] in agreement with our inference from the measured $C_b$, and does explain why no structural transformation has been observed by determining the Hugoniot curve up to 560 GPa,[33] which is insensitive to phase transitions with slight or no volume change.

Although the first interpretation gains supports from both experiment and theory, we still cannot exclude one other possible explanation of the observed discontinuity at ~60 GPa, that is, a pressure-induced electronic topological transition. Recently, Antonangeli *et al.* observed softening of the shear sound velocity around 90 GPa by using the inelastic X-ray scattering technique and DAC without pressure-transmitting medium at room temperature.[34] Supported by first-principles density-functional theory,[34-36] they suggest that the shear softening results from *s-d* electronic transitions, although a structural transformation does not ensue due to the limited energy gain associated with such transitions. If we take shock-induced temperature increments (~1000 K at 60 GPa) into account, the onset of shear softening could shift down to lower pressures, thereby becoming more consistent with our observation.

## IV. SUMMARY

Within the limits of our optical measurement technique, we cannot exactly ascribe the observed longitudinal sound velocity discontinuity at shock pressure 60 GPa to either a structural transformation or an electronic topological transition. However, either one will shed light on understanding the existing discrepancy in the high-pressure melting curves of Ta. The former could imply that the respective melting curves obtained from the LHDAC and SW experiments actually represent different melting paths,[25] one from the bcc phase and another from the shock-induced phase. The latter is also expected to have a strong influence upon melting from electronic rearrangements.[15] Ultrafast time-resolved X-ray diffraction could offer a powerful *in situ* technique to detect shock-induced structural changes and to clarify this confusion.

## ACKNOWLEDGMENTS

We are grateful to Z. Song and K. Jin for helpful discussions and to D. Alfè for critical reading, and also thankful to Y. Xiang, H. Chen, and J. Li for their assistance in performing experiments. This work was jointly supported by the National Natural Science Foundation of China under Grant No. 11064007 and the Foundation of Laboratory for Shock Wave and Detonation Physics Research, CAEP under Grant No. 9140C6702030802. Z.L. acknowledges support from the Program for New Century Excellent Talents in University.

a) Present address: Materials and Structures Laboratory, Tokyo Institute of Technology, R3-10, 4259 Nagatsuta, Yokohama 226-8503, Japan, Email: jianbohu@caep.ac.cn

b) Email: cddai@caep.ac.cn

**List of figure captions**

**Fig.1**(color online) Schematic of the experimental configuration for sound velocity measurements using the direct reverse-impact technique and velocity interferometry.

**Fig.2**(color online) Three selected particle velocity profiles of Ta observed at Ta/LiF interface at various shock pressures. The velocity profiles are displaced along the time axis to clearly show the details of the shock front. The inset shows an amplification of the elastic precursor in the velocity profile at 12.7 GPa.

**Fig.3** (color online) Euler longitudinal and bulk sound velocities of Ta plotted against shock pressure. Solid line is the calculated bulk sound velocity using Grüneisen equation of state (Ref.[21]). Dashed and dash-dot lines are, respectively, the calculated longitudinal sound velocities of bcc and unknown phase (bcc or other structure). The parameters used to calculate the sound velocities are listed as follows (Refs.[18] and [2]): $\rho_0$=16.65 g/cm$^3$, $C_0$=3.329 km/s, $\lambda$=1.307, $\gamma_0$=1.8, $\upsilon$=0.34 for the bcc phase, and $\upsilon$=0.41 for the unknown phase deduced from the sound velocity data. Here $C_0$ and $\lambda$ are the Hugoniot parameters, and $\upsilon$ the Poisson ratio.

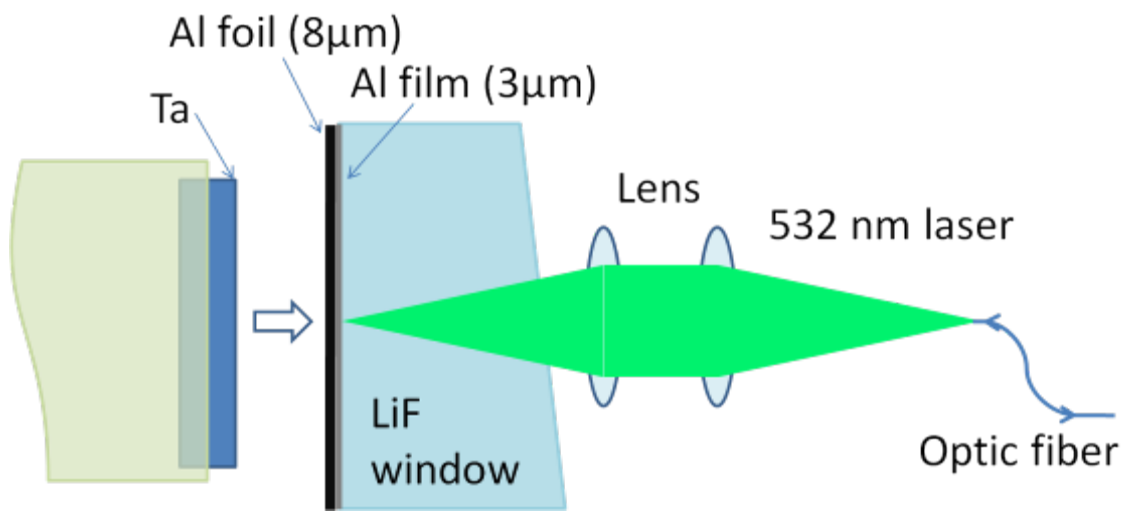

**Fig.1 Hu** *et al.*

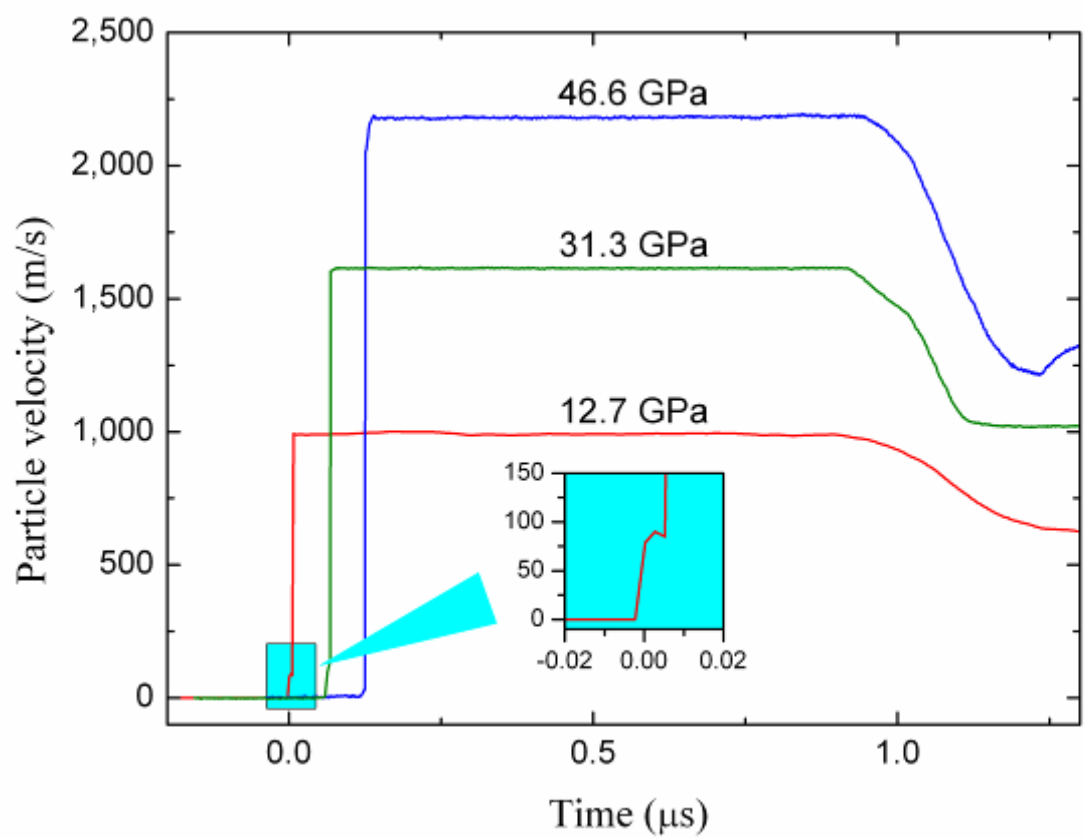

**Fig.2 Hu** *et al.*

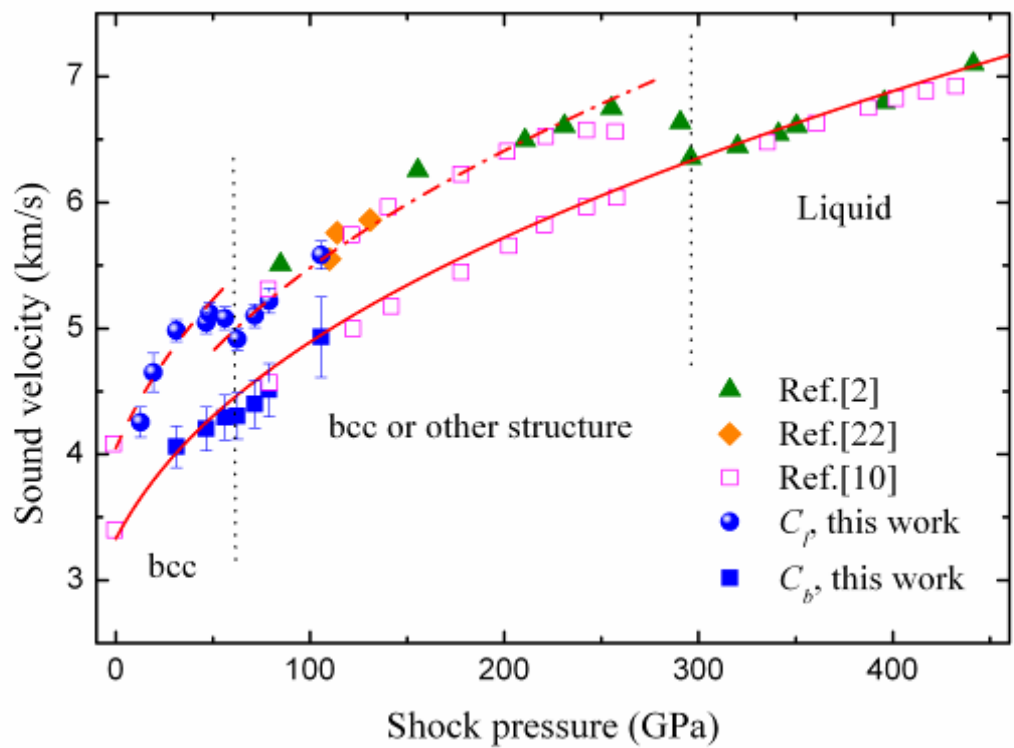

**Fig.3 Hu *et al*.**